# LEARNING COMPACT STRUCTURAL REPRESENTATIONS FOR AUDIO EVENTS USING REGRESSOR BANKS

*Huy Phan*⋆†, *Marco Maass*⋆, *Lars Hertel*⋆, *Radoslaw Mazur*⋆, *Ian McLoughlin*‡, and *Alfred Mertins*⋆

⋆Institute for Signal Processing, University of Lübeck, Germany
†Graduate School for Computing in Medicine and Life Sciences, University of Lübeck, Germany
‡School of Computing, University of Kent, UK
{phan, maass, hertel, mazur, mertins}@isip.uni-luebeck.de, ivm@kent.ac.uk

## ABSTRACT

We introduce a new learned descriptor for audio signals which is efficient for event representation. The entries of the descriptor are produced by evaluating a set of regressors on the input signal. The regressors are class-specific and trained using the random regression forests framework. Given an input signal, each regressor estimates the onset and offset positions of the target event. The estimation confidence scores output by a regressor are then used to quantify how the target event aligns with the temporal structure of the corresponding category. Our proposed descriptor has two advantages. First, it is compact, i.e. the dimensionality of the descriptor is equal to the number of event classes. Second, we show that even simple *linear* classification models, trained on our descriptor, yield better accuracies on audio event classification task than not only the nonlinear baselines but also the state-of-the-art results.

*Index Terms*— feature learning, audio event, recognition, structural encoding

## 1. INTRODUCTION

Nonspeech acoustic event detection and classification are important problems of computational auditory scene analysis [1] and have recently received great attention [2, 3]. However, compared to speech recognition, which is mature, these problems are still in their infancy.

In general, the characteristics of audio events differ from those of speech in the manner that they cover a much wider variety in frequency content and duration. Nevertheless, they are similar in one perspective. Speech exposes temporal structure, i.e. it is possible to decompose words into their constituent phonemes. Likewise, an audio event can be decomposed into atomic units of sound [4, 5]. For example, the sound of "using water tap" event may further be decomposed into the sounds of the water running in the faucet, then pushing into the air, and finally splashing into the sink. As a result, the patterns of unit sequences can be used as a signature to distinguish different sound events. However, unlike phonemes in speech, it is not clear how to design or discover the sound unit dictionary to encode all sound events. Furthermore, comparing sequential patterns under high intra- and inter-category variations of audio events is not a trivial task.

Regarding audio event representation, besides widely used hand-crafted descriptors such as mel-scale filter banks [6], log frequency filter banks [7], and time-frequency features [8], learned descriptors are becoming more and more common. They range from codebook-based representations (e.g. bag-of-words models [9, 10], sparse coding [11], non-negative matrix factorization [12], and examplar-based coding [13]), to sharing features [14], deep features [15, 16], and speech-based generic features [17]. Recently, structural information from audio events has been shown essential for the recognition task [10, 18]. There have been several attempts to capture these types of information. An audio event can be considered as a sequence of atomic units of sound [4, 19] and the pattern of occurrences is then used as an event signature. Alternatively, the temporal configuration can also be encoded using self-organizing maps [18], pyramid bag-of-words models [10], and audio phrases [20].

In this work, we aim to quantify the temporal structure of an event category as a scalar, which allows one to justify how an event aligns to the temporal configuration of an event category. For instance, how much a "door knock" instance aligns with the temporal structure of "chair moving" class. Furthermore, it facilitates numerical measurements of the similarity between events, such as for the classification task. We propose to address this by employing a regression model which takes the audio signal as input and provides estimates for the event onset and offset positions as probability density functions. Instead of using the regressor as in [21, 22], we consider the predicted confidence scores of the regressor as structural measures. Further, we propose a global descriptor which is obtained by evaluating a bank of class-specific regressors on the target event. The responses of the regressor bank quantify the alignment of the event to the structures of different event classes, and hence, encode the shared features between different classes.

The proposed descriptor is compact since the number of entries equals the number of event classes. Furthermore, using the proposed descriptor, we obtain state-of-the-art accuracy on event classification even with simple *linear* classification models. The intuition behind our proposed descriptor is that it provides a semantically rich representation of an audio event by measuring how it aligns to the structure of different event classes. Thus, a linear classifier trained on this representation will express an event class as a linear combination of the structural alignments.

## 2. LEARNING DECISION FORESTS REGRESSORS

### 2.1. Training

In this section, we describe how to learn a class-specific regression model for event onset and offset estimation. The regressor is learned as in [21, 23] using the random decision forests framework [24]. We first decompose the training audio signals to obtain the set of audio segments $\mathcal{S} : \{s_n = [\mathbf{x}_n, \mathbf{d}_n]; n = 1 \ldots N_\mathcal{S}\}$, where $\mathbf{x}_n \in \mathbb{R}^M$ is the feature vector for the segment $n$, and $M$ is the dimensionality.

This work was supported by the Graduate School for Computing in Medicine and Life Sciences funded by Germany's Excellence Initiative [DFG GSC 235/1].

$\mathbf{d}_n = [d_n^+, d_n^-] \in \mathbb{R}_+^2$ is the distance vector where $d_n^+$ and $d_n^-$ denote the distances from $n$ to the first segment (e.g. the onset) and the last segment (e.g. the offset) inclusive of the corresponding event respectively.

The forest model consists of $T$ binary decision trees. Each tree is constructed as follows. We randomly sample and use a subset of audio segments from $\mathcal{S}$. Starting from the root node with the full set of audio segments, we randomly generate a pool of binary tests. We then choose an optimal test to divide the data set into two subsets, each is assigned to one of the child nodes. This splitting process is repeated recursively to grow the tree. The growing stops at a leaf node when either a maximum depth $D_{\max}$ of the tree is reached or a minimum number $N_{\min}$ of audio segments are left. At a leaf node, the distances to the event onset and offset of its audio segments are modeled. In this manner, the entire forest is constructed.

The binary tests at the split nodes are given by

$$t_{r,\tau}(\mathbf{x}) = \begin{cases} 1, & \text{if } \mathbf{x}^r > \tau \\ 0, & \text{otherwise.} \end{cases} \quad (1)$$

Here, $\mathbf{x}^r$ denotes the value of $\mathbf{x}$ at a random selected feature channel $r \in \{1, \ldots, M\}$. $\tau$ is a random threshold generated in the range of $\mathbf{x}^r$. Hereafter, the best test is adopted to minimize the total distance variation:

$$t_{r,\tau}^* = \underset{t_{r,\tau}}{\operatorname{argmin}} \big( \sum_i \|\mathbf{d}_i^{\text{left}} - \bar{\mathbf{d}}^{\text{left}}\|_2^2 + \sum_i \|\mathbf{d}_i^{\text{right}} - \bar{\mathbf{d}}^{\text{right}}\|_2^2 \big), \quad (2)$$

where $\bar{\mathbf{d}}$ denotes the mean distance vector of the corresponding subset indicated by the superscript.

The onset and offset distances of the audio segments that arrived at a leaf are modeled as Gaussian distributions $\mathcal{N}^+(d; \bar{d}^+, \Sigma^+)$ and $\mathcal{N}^-(d; \bar{d}^-, \Sigma^-)$, respectively. Here, $(\bar{d}^+, \Sigma^+)$ and $(\bar{d}^-, \Sigma^-)$ denote the means of variances of onset and offset distances of the audio segments.

### 2.2. Event onset and offset estimation

Given a test audio segment $\mathbf{x}_{n'}$ at the time index $n'$, we want to estimate where the onset and offset should be. Input $\mathbf{x}_{n'}$ into a tree, at every split node, the stored binary question is evaluated, directing $\mathbf{x}_{n'}$ to the right or left child until ending up at a leaf node. The estimates for the onset and offset positions are then given by:

$$p^+(n|\mathbf{x}_{n'}, \bar{d}^+, \Sigma^+) = \mathcal{N}^+(n; n' - \bar{d}^+, \Sigma^+), \quad (3)$$

$$p^-(n|\mathbf{x}_{n'}, \bar{d}^-, \Sigma^-) = \mathcal{N}^-(n; n' + \bar{d}^-, \Sigma^-). \quad (4)$$

The interpretation for (3) and (4) is that we shift the Gaussian distributions $\mathcal{N}^+(d; \bar{d}^+, \Sigma^+)$ at $\bar{d}^+$ backward from $n'$ and $\mathcal{N}^-(d; \bar{d}^-, \Sigma^-)$ at $\bar{d}^-$ forward from $n'$. The estimation by the forest is computed by averaging over all trees:

$$p^+(n|\mathbf{x}_{n'}) = \frac{1}{T} \sum_{t=1}^{T} p^+(n|\mathbf{x}_{n'}, \bar{d}_t^+, \Sigma_t^+), \quad (5)$$

$$p^-(n|\mathbf{x}_{n'}) = \frac{1}{T} \sum_{t=1}^{T} p^-(n|\mathbf{x}_{n'}, \bar{d}_t^-, \Sigma_t^-). \quad (6)$$

## 3. LEARNING COMPACT STRUCTURAL DESCRIPTORS WITH A BANK OF REGRESSORS

### 3.1. Regressors for structural measurements

Suppose that we have learned a regressor $\mathcal{R}_c$ for a target event class $c$ as in Section 2. Given an audio signal, we decompose it into a sequence of segments $(\mathbf{x}_n; n = 1 \ldots N)$. The regressor takes an audio segment $\mathbf{x}$ as input and makes estimates for the event onset and offset positions in terms of the probability density functions $p^+(n|\mathbf{x}, c)$ and $p^-(n|\mathbf{x}, c)$ given in (5) and (6), respectively. The respective onset and offset estimation confidence scores $f_{c,\mathbf{x}}^+(n)$ and $f_{c,\mathbf{x}}^-(n)$ are then computed by

$$f_{c,\mathbf{x}}^+(n) = p^+(n, c|\mathbf{x}) = P(c|\mathbf{x})p^+(n|\mathbf{x}, c), \quad (7)$$

$$f_{c,\mathbf{x}}^-(n) = p^-(n, c|\mathbf{x}) = P(c|\mathbf{x})p^-(n|\mathbf{x}, c). \quad (8)$$

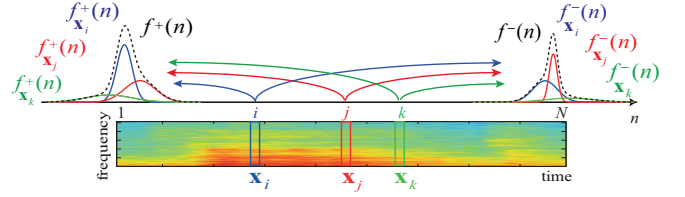

**Fig. 1**. **Illustration of event onset and offset estimation.** The total estimation confidence scores $f^+(n)$ and $f^-(n)$ for the target event onset and offset are computed by summing the individual scores obtained by three segments $\mathbf{x}_i$, $\mathbf{x}_j$, and $\mathbf{x}_k$. We ignored the class label here for simplicity.

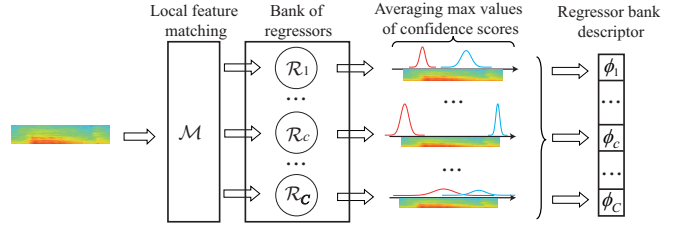

**Fig. 2**. **Regressor bank for feature extraction.** The local feature matching is performed by the multi-class classifier $\mathcal{M}$. A class-specific regressor $\mathcal{R}_c$ is trained using training samples of class $c$. The bank of these regressors is then used as feature extractors to produce a semantically rich descriptor of the input audio event.

Here, $P(c|\mathbf{x})$ is the probability that the local feature $\mathbf{x}$ matches to event class $c$. The estimates for all segments read

$$f_c^+(n) = \sum_{i=1}^{N} f_{c,\mathbf{x}_i}^+(n), \quad (9)$$

$$f_c^-(n) = \sum_{i=1}^{N} f_{c,\mathbf{x}_i}^-(n). \quad (10)$$

The estimation is illustrated in Fig. 1. Via the learned regression model, the local features vote for the boundary of the audio event. By this, we implicitly model the "shape" of the audio event, e.g. its temporal extent, as a constellation of local features. Since we intend to estimate the onset and offset positions separately, the regression confidence scores can be reasonably considered as the measures for forward and backward structures of the event.

Finally, we average the maximum values of the confidence scores to produce the structural descriptor $\phi_c$ of the event class $c$ measured on the input audio event:

$$\phi_c = \frac{1}{2} \big( \max_n \big( f_c^+(n) \big) + \max_n \big( f_c^-(n) \big) \big). \quad (11)$$

### 3.2. Regressor bank descriptor

In general, the confidence scores can be compared against a threshold for recognition. Good performance has been demonstrated in many applications [22, 24, 21, 23]. However, simple thresholding will ignore the sharing of features between different classes which is important to boost the performance [25, 14].

We therefore propose stacking the regressors in a bank as in Fig. 2. The regressors then play the role of a mid-level feature extractor to produce an intermediate representation for the audio event. In other words, we transform a sequence of audio segments $(\mathbf{x}_n; n = 1 \ldots N)$ of the input signal into a compact descriptor $\boldsymbol{\phi} = [\phi_1, \ldots, \phi_C]^T \in \mathbb{R}_+^C$. Here, $C$ is the number of event categories of interest and $\boldsymbol{\phi}$ is given in (11). As a result, the audio event is embedded in the space spanned by the responses of the regressors.

The descriptor can be interpreted as how the audio event aligns to the temporal configurations of different event classes modeled by the regressors. Since the regressor bank features are semantically rich representations, even simple linear classification models trained on our descriptor achieve good classification accuracies.

## 4. EXPERIMENTS

### 4.1. Experiment setup

**Parameters.** The audio signals were downsampled to 16 kHz. Each audio event was decomposed into a sequence of 50 ms segments with 10 ms overlap. We also conducted experiments with various segment sizes $(30, 40, \ldots, 100$ ms$)$ to see how the performance changes.

We trained the regression models with ten trees each. We randomly sampled 50% of the training set to learn a tree. Furthermore, we set the number of binary tests generated at a split node to 20,000, the maximum depth to $D_{\max} = 12$, and the minimum number of audio segments for early termination to $N_{\min} = 20$. For local feature matching, we trained a classifier $\mathcal{M}$ using random forests [26] with 200 trees. For the purpose of classification, an audio segment was labeled with the label of the event from which it originated.

**Low-level features to represent audio segments.** Although almost any arbitrary features can be used to describe an audio segment, we used a set of very basic acoustic features: 16 log-frequency filter bank coefficients [7], their first and second derivatives, zero-crossing rate, short-time energy, four sub-band energies, spectral centroid, and spectral bandwidth. The overall feature dimension is 53.

**Classification models.** Our final classification systems were trained on the bank-of-regressors (BoR) descriptors extracted from the training audio events using one-vs-one standard SVM (BoR-linear) and $\chi^2$-kernel SVM (BoR-$\chi^2$). To extract the descriptors for the training events, we conducted 10-fold cross validation on the training data. We noticed that it is unnecessary to conduct cross-validation on the regression forests but can simply employ those trained with the whole training data. It turned out that we only need to do cross-validation for the classifier $\mathcal{M}$ used for feature matching.

An entry $\phi_c$ of a descriptor was normalized to $\frac{\phi_c}{\max_{\phi_c}}$, where $\max_{\phi_c}$ is the maximum value of $\phi_c$ in the training events. The descriptors are further $\ell_1$-norm normalized. The hyperparameters of the SVMs were tuned via leave-one-out cross-validation.

**Baseline systems.** We compare performance of our systems with three baseline systems:

1. *Bag-of-words system (BoW)*. We implemented a BoW model which has been widely used for audio event recognition [9, 27, 10]. Using this model, an audio event is represented by a histogram of codebook entries.
2. *Pyramid bag-of-words system (PBoW)*. We extracted BoW descriptors on different pyramid levels [28] to encode temporal structure of the audio events. This approach has recently achieved state-of-the-art results on different benchmark datasets [10].
3. *Max voting system*. This system assigns a test audio event the class label that corresponds to the regressor with maximum response in the bank. It is equivalent to the one proposed in [21] if we consider it for classification purpose only.

For the BoW and PBoW baselines, we used $k$-means for codebook learning. The entries were obtained as the cluster centroids, and codebook matching was based on Euclidean distance. We used different codebook sizes $(50, 75, \ldots, 250)$. In particular, we tried 2, 3, and 4 pyramid levels for the PBoW systems. In addition to standard SVM, nonlinear SVMs with RBF, $\chi^2$, and histogram intersection (*hist.* for short) kernels were also implemented. All the hyperparameters were tuned by cross-validation. Finally, the systems with the best performance were compared with our systems.

**Datasets.** We tested our approach on the four following datasets with different degrees of complexity:

1. *ITC-Irst* [31]. It consists of 741 audio events of 16 categories. We evaluated on twelve categories and used nine first recording sessions for training and three remaining sessions for testing as publicly available settings [31, 14]. Only single-channel data named *TABLE_1* was used.
2. *UPC-TALP* [32]. For this dataset, we used a single channel (*channel 10*) of 8 recordings with isolated events. There are 1,418 instances of eleven categories. Following [29], we alternatively used seven sessions for training and the remaining session for testing. The average accuracy is finally reported.
3. *Freiburg-106* [33]. There are 1,479 audio human activities of 22 categories. As in [33, 20], the test set contains every second recording of a class, and the training set contains all the remaining recordings.
4. *NAR* [30]. Overall, it consists of 852 sound signals of 42 classes. Particularly, it includes some speech categories, and they are also treated as audio events in general. As in [30], we randomly split the data into ten parts and conducted 10-fold cross-validation. The average accuracy is then reported.

### 4.2. Experimental results

**Responses of regressor bank.** For illustration, we show in Fig. 3 the normalized responses of the regressor bank on typical examples of different categories of the ITC-Irst dataset. Note that we padded zeros to the beginning and end of each sequence to make it five times longer before regression to account for event duration variations. It can be seen that some examples (e.g. "paper wrapping", "phone ring") are very discriminative, so that the max voting scheme should be adequate for recognition. However, it would yield wrong recognition on many other classes (e.g. "door slam", "key jingle"). Our model overcomes this difficulty by effectively discriminating on combinations of shared features.

**Performance comparison.** The classification performances achieved by different systems are summarized in Table 1. For the BoW and PBoW baselines, the $\chi^2$ and hist. kernels were found most appropriate. This is expected since they are based on histogram representations. On the other hand, a pyramid level of two is optimal for the PBoW baselines. However, it is worth emphasizing that performances of the baselines are not consistent. For instance, on the ITC-Irst dataset, the best BoW is with codebook size of 50 while it is 200 for the Freiburg-106 dataset. Similarly, those for PBoW are 225 and 200, respectively. It is also noticed that BoW performs better than PBoW on the ITC-Irst and UPC-TALP datasets, but the opposite results are seen on the Freiburg-106 and NAR datasets. On another hand, as expected, by taking into account the sharing of features between different classes, our systems significantly boost the classification performance to a higher level compared to the simple maximum voting strategy. We also list the best performance previously reported for those datasets.

From Table 1, it can be seen that our BoR-$\chi^2$ systems consistently outperform all baselines and state-of-the-art systems on three out of four datasets. Overall, the improvements over the best competitors (e.g. the baselines or the state-of-the-art systems) are 0.6%, 0.1%, and 0.6% with respect to the ITC-Irst, UPC-TALP, and NAR datasets. Furthermore, even with simple linear classifiers, e.g. the BoR-linear systems, we are able to achieve better performance than the non-linear baselines on the ITC-Irst, Freiburg-106, and NAR datasets. We would like to point out that the standard SVMs are very efficient to train and evaluate compared to nonlinear baseline models. On individual classification accuracies, our BoR-$\chi^2$ systems achieve equivalent or higher accuracies on 10 out of 12, 7 out of 11, 20 out of 22, and 33 out of 42 categories compared to all baselines for ITC-Irst, UPC-TALP, Freiburg-106, and NAR datasets, respectively. The single exception is the Freiburg-106 test, where our previously published method [20] is able to perform very slightly better than the current system. It is worth pointing out that the BoR descriptors are significantly more compact than the high-dimensional audio phrases used in [20].

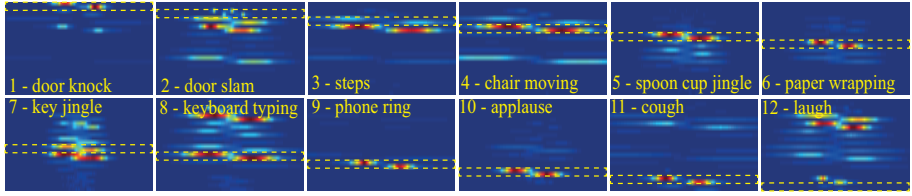

**Fig. 3**. **ITC-Irst dataset.** Responses of regressor bank on audio events of different classes. Each plot is associated with the class identity numbers and the class names. For an event of class $c$, the responses of the regressor $\mathcal{R}_c$ are located in the dash-line boxes, the onset score on one row followed by the offset score on the other row.

**Table 1**. **Overall classification performance.** Comparison of classification accuracies obtained by different systems. The results on the Freiburg-106 dataset are reported on *f-score* (%) to consent with the work in [20].

| Dataset | BoW | PBoW | Max voting | Best reported | Our systems | | |
|---|---|---|---|---|---|---|---|
| | | | | | BoR-linear | BoR-$\chi^2$ | BoR+ |
| ITC-Irst | 97.3 | 96.6 | 95.9 | 97.3 [14] | **97.9** | 97.9 | 99.3 |
| UPC-TALP | 96.6 | 96.5 | 94.5 | 87.6 [29] | 95.8 | **96.7** | 96.8 |
| Freiburg-106 | 96.6 | 96.8 | 92.3 | **98.9** [20] | 97.2 | 97.8 | 98.1 |
| NAR | 94.8 | 96.4 | 92.6 | 97.0 [30] | 96.8 | **97.6** | 97.6 |

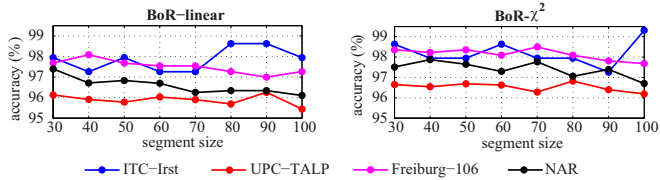

**Fig. 4**. **Variation of segment size.** Classification accuracies as functions of audio segment size.

**Varying the size of audio segments.** In this experiment, we studied how the classification accuracy of the proposed method changes as a function of the audio segment size. We performed this analysis using different sizes in the set $(30, 40, \ldots, 100\text{ ms})$ and reported the results in Fig. 4. One can see that the performance is quite stable, except for the ITC-Irst case. This can be explained by its small number of test samples. On average, our BoR-$\chi^2$ systems obtain $98.2\pm0.6\%$, $96.5\pm0.2\%$, $98.1\pm0.3\%$, and $97.4\pm0.4\%$ on ITC-Irst, UPC-TALP, Freiburg-106, and NAR datasets, respectively.

**Combination with unstructured features.** Our BoR descriptor works well for event categories which clearly expose structure. On the contrary, it appears to be slightly worse on weakly structured events, such as those with impulse-like signals (e.g. "door slam" in the ITC-Irst dataset). For those kinds of events, unstructured features (like bag-of-words) tend to yield better results. It is therefore reasonable to somehow combine our structural descriptor with an unstructured descriptor. For example, our experiment shows improvements of 0.7% and 0.2% on the ITC-Irst and Freiburg-106 accuracies, respectively, compared to the BoR-$\chi^2$ system when combining the BoR descriptor with the bag-of-words features in the BoW baselines. However, it is quite costly to build a second system for combination. Alternatively, we exploit the random forest classifier $\mathcal{M}$ used for feature matching to form a compact unstructured descriptor at very little extra computational cost.

For an event decomposed into a sequence of $N$ audio segments $(\mathbf{x}_n; n = 1, \ldots, N)$, we obtained an unstructured descriptor denoted as $\varphi = [\varphi_1, \ldots, \varphi_C]^T \in \mathbb{R}_+^C$ where

$$\varphi_c = \frac{1}{N} \sum_{n=1}^{N} P(c|\mathbf{x}_n). \quad (12)$$

The vector $\varphi$ is then $\ell_1$-norm normalized. Different descriptors are combined using an extended Gaussian kernel [34]:

$$K(e_i, e_j) = \exp\big(-\sum_{k \in \{\phi, \varphi\}} \frac{1}{A^k} D(e_i^k, e_j^k)\big), \quad (13)$$

where $D(e_i^k, e_j^k)$ is the $\chi^2$ distance between the audio events $e_i$ and $e_j$ with respect to the $k$-th channel. $A^k$ is the mean value of the $\chi^2$ distances between the training samples for the $k$-th channel. For classification, we learned a nonlinear SVM with the kernel $K$ defined in (13). It leads to improvements of 1.4%, 0.1%, 0.3% on the ITC-Irst, UPC-TALP, and Freiburg-106 datasets, respectively, as shown for the BoR+ systems in Table 1. This simple fusion approach could in future be augmented with methods such as multiple kernel learning frameworks.

## 5. CONCLUSIONS

We have presented a new structural descriptor for efficient audio event classification. It focuses on the learning of a mid-level representation that enables us to measure how an audio event lines up with temporal structures of different event categories of interest. The temporal structures are modeled by class-specific regressors which are based on a random decision forest framework. Our mid-level features are produced by evaluating a set of pre-trained regressors over the input audio event. Experiments on four benchmark datasets show the efficiency of our descriptor in terms of classification accuracy.